\documentclass[twocolumn,amsmath]{aastex702}

\newcommand{\ha}{H$\alpha$}

\newcommand{\w}{$W_r$~}

\newcommand{\z}{$z$}

\newcommand{\ctwo}{[C~{\sc ii}]}

\newcommand{\sii}{[S~{\sc ii}]}
\newcommand{\nii}{[N~{\sc ii}]}
\revised{\today}
\submitjournal{AAS}
\shorttitle{Galaxy Overdensities at $z\sim5.2$ }
\shortauthors{Geng et al.}
\begin{document}
\title{\boldmath JWST Reveals Contrast Overdensities around Two ALMA-CRISTAL Galaxies at $z\sim5.2$}
\author[0009-0009-4187-2095]{Shaoze Geng}
\affiliation{Chinese Academy of Sciences South America Center for Astronomy, National Astronomical Observatories, CAS, Beijing 100101, China}
\affiliation{School of Astronomy and Space Sciences, University of Chinese Academy of Sciences, Beijing 100049, China}
\email{}

\correspondingauthor{Siwei Zou}
\author[0000-0002-3983-6484]{Siwei Zou}
\affiliation{Chinese Academy of Sciences South America Center for Astronomy, National Astronomical Observatories, CAS, Beijing 100101, China}
\affiliation{Departamento de Astronom\'ia, Universidad de Chile, Casilla 36-D, Santiago, Chile}
\email{zousw@bao.ac.cn}

\author[0000-0002-6290-3198]{Manuel Aravena}
\affiliation{Instituto de Estudios Astrof\'isicos, Facultad de Ingenier\'ia y Ciencias, Universidad Diego Portales, Av. Ej\'ercito 441, Santiago, Chile}
\affiliation{Millennium Nucleus for Galaxies (MINGAL)}
\email{}

\author[0000-0001-5847-7934]{Nikolaus Z. Prusinski}
\affiliation{Chinese Academy of Sciences South America Center for Astronomy, National Astronomical Observatories, CAS, Beijing 100101, China}
\affiliation{Departamento de Astronom\'ia, Universidad de Chile, Casilla 36-D, Santiago, Chile}
\email{}

\author[0000-0002-6510-5028]{Yi W. Ren}
\affiliation{Chinese Academy of Sciences South America Center for Astronomy, National Astronomical Observatories, CAS, Beijing 100101, China}
\affiliation{Departamento de Astronom\'ia, Universidad de Chile, Casilla 36-D, Santiago, Chile}
\email{}

\author[0000-0002-9382-9832]{Andreas L. Faisst}
\affiliation{IPAC, California Institute of Technology, 1200 E. California Blvd. Pasadena, CA 91125, USA}
\email{}

\author[0000-0001-7457-4371]{Lilian~L.~Lee}
\affiliation{Max-Planck-Institut f\"ur Extraterrestrische Physik (MPE), Giessenbachstr. 1, D-85748 Garching, Germany}
\email{}

\author[0009-0004-1270-2373]{Lun-Jun Liu}
\affiliation{Department of Physics, Washington University in St. Louis, One Brookings Drive, St. Louis, MO, 63130 USA}
\email{}

\author[0000-0002-8136-8127]{Juan Molina}
\affiliation{Instituto de F\'isica y Astronom\'ia, Universidad de Valpara\'iso, Avda. Gran Breta\~na 1111, Valpara\'iso, Chile}
\affiliation{Millennium Nucleus for Galaxies (MINGAL)}
\email{}

\author[0000-0002-9948-3916]{Michael Romano}
\affiliation{Max-Planck-Institut für Radioastronomie, Auf dem Hügel 69, 53121 Bonn, Germany}
\affiliation{INAF - Osservatorio Astronomico di Padova, Vicolo dell’Osservatorio 5, I-35122 Padova, Italy}
\email{}

\author[0009-0003-0031-0527]{Bahareh S. Salmasi}
\affiliation{School of Mathematical and Physical Sciences, Macquarie University, Sydney, NSW 2109, Australia}
\affiliation{Astrophysics and Space Technologies Research Centre, Macquarie University, Sydney, NSW 2109, Australia}
\email{}

\begin{abstract}
The distribution of stellar mass between massive galaxies and their companions traces how galaxy assembly proceeds within early large-scale structure. We investigate the environments of two galaxies from the Atacama Large Millimeter/submillimeter Array (ALMA) CRISTAL survey: CRISTAL-19 (C19; $z=5.235$) and CRISTAL-21 (C21; $z=5.256$), using F444W slitless spectroscopy from the Near-Infrared Camera (NIRCam) on the James Webb Space Telescope (JWST) in COSMOS-3D. A friends-of-friends analysis of \ha~emitters identifies two distinct overdense nodes separated by approximately 25~comoving Mpc. Although C21 is the more massive galaxy, C19 has the more massive, actively star-forming companions. The companion populations around C19 and C21 have median stellar masses of $\log(M_\star/M_\odot)=9.43$ and 8.82 and median dust-corrected \ha~star formation rates of 31 and $16~M_\odot~\mathrm{yr}^{-1}$, respectively. Their star formation rate contrast largely reflects the difference in companion masses. C19 contains only $4.5^{+0.9}_{-0.7}\%$ of the combined measured stellar mass of itself and its selected companions; the corresponding fraction for C21 is $39^{+5}_{-5}\%$. Thus, stellar mass remains more widely distributed among galaxies around C19, whereas a larger share resides in C21. Resolved interacting companions around C19 and \ctwo~kinematics consistent with a late-stage merger in C21 support this contrast in assembly state. These findings support the phase-dependent picture of \citet{Zou2026}, in which the surrounding star-forming population changes as interactions and mergers build up a massive galaxy.
\end{abstract}


\section{Introduction}

The assembly of massive galaxies and the growth of large-scale structure are linked through gas accretion and mergers in the early cosmic web \citep{Dekel2009,Overzier2016}. Star formation also produces the heavy elements needed for dust formation and grain growth \citep{Schneider2024}. Substantial dust reservoirs in galaxies at $z\gtrsim4$ show that obscured stellar growth was already underway within the first few billion years \citep{Mitsuhashi2024,Inami2022}. How this growth is distributed across assembling structures remains uncertain. A massive galaxy may have accumulated a large stellar mass while substantial growth continues in its neighbours. Measuring both populations is necessary to connect the assembly of an individual galaxy to that of its environment.

Spatially resolved observations now map stars and the gas and dust of the interstellar medium (ISM) in early galaxies. The \ctwo~Resolved ISM in STar-forming galaxies with ALMA (CRISTAL) survey is a Cycle~8 Large Program of the Atacama Large Millimeter/submillimeter Array (ALMA) targeting star-forming galaxies at $4\lesssim z\lesssim6$ \citep{HerreraCamus2025}. It reveals rotating disks, clumpy systems, mergers, and extended \ctwo~emission \citep{HerreraCamus2025,Lee2025}. 
The ALPINE--CRISTAL--JWST survey combines observations from the ALMA Large Program to INvestigate \ctwo~at Early times (ALPINE) and CRISTAL with JWST/NIRSpec integral field unit (IFU) spectroscopy and HST/JWST imaging to map star formation, dust attenuation, and chemical enrichment on kiloparsec scales \citep{Faisst2026,Fujimoto2025}. These data characterize dust and the interstellar medium and probe gas interactions with its small scale circumgalactic medium, but do not sample their wider environments. NIRCam wide-field slitless spectroscopy (WFSS) is thus ideally suited to mapping the large-scale environments of massive galaxies.

COSMOS-3D (hereafter C3D; GO\#5893; P.I. K.~Kakiichi) obtained F444W WFSS observations over $0.33~\mathrm{deg}^{2}$ in the Cosmic Evolution Survey (COSMOS) field, within the COSMOS-Web footprint (GO\#1727; P.I. J.~Kartaltepe and C.~M.~Casey). Using these data, \citet{Zou2026} studied 18 massive and dusty star-forming galaxies at $z=4.9$--7.2, including six ALMA-CRISTAL galaxies. Interacting systems show the strongest compact excesses of \ha~emitters (HAEs), whereas MIRI-bright hot-dust or obscured active galactic nucleus (AGN) candidates show weaker companion excesses. This contrast motivated a phase-dependent picture linking interactions, dusty growth, and coalescence to the visible companion population. We test this picture beyond companion counts using stellar masses and star formation rates (SFRs) to determine where mass has accumulated and where growth continues within the surrounding populations.

We focus on CRISTAL-19 at $z=5.235$ (C19; also known as DEIMOS\_COSMOS\_494763) and CRISTAL-21 at $z=5.256$ (C21; also known as HZ7), which show the strongest \ha-emitter overdensities among the six CRISTAL galaxies in \citet{Zou2026}. We identify their spectroscopic groups and compare the masses, SFRs, and morphologies of their companions to determine how stellar growth is distributed between each target and its surrounding population.

This paper is structured as follows. Section~\ref{sec:data_reduction} describes the data reduction, photometry, emission-line selection, and previously reported properties of C19 and C21. Section~\ref{sec:results} presents the companion samples, overdensities, stellar masses, and SFRs. Section~\ref{sec:discussion} discusses the contrasting assembly states and possible descendants of the two nodes, and Section~\ref{sec:summary} summarizes our findings.

We adopt a flat cosmology containing a cosmological constant and cold dark matter ($\Lambda$CDM), with $H_0=70~\mathrm{km~s^{-1}~Mpc^{-1}}$ and $\Omega_{\mathrm{m}}=0.3$.

\section{Data Reduction and Analysis}
\label{sec:data_reduction}

\subsection{NIRCam WFSS Data Reduction}
\label{sec:wfss_reduction}

We use COSMOS-Web F444W imaging \citep{Casey2023,Franco2026} and C3D F444W/GRISMR spectroscopy. The grism observations cover approximately $60\%$ of the COSMOS-Web NIRCam imaging area, span $3.9$--$5.0~\mu\mathrm{m}$, and have resolving power $R\simeq1600$ at $4~\mu\mathrm{m}$. The representative $5\sigma$ line-flux limit is $4\times10^{-18}~\mathrm{erg~s^{-1}~cm^{-2}}$ near $4.5~\mu\mathrm{m}$.

The reduction follows \citet{Sun2023}. We processed exposures through Stage~1 of the JWST pipeline, applied filter- and detector-matched imaging flats, and subtracted sky backgrounds estimated from sigma-clipped median grism frames. Commissioning-based, field-dependent models provide the spectral traces and wavelength calibration; observations of P330-E provide the flux calibration. We combined the calibrated two-dimensional spectra for the line search. The \ha~spectra of all selected companions around C19 and C21 are presented in Figure~\ref{fig:c19_c21}(c). The JWST/NIRCam imaging shown in Figure~\ref{fig:c19_c21} was retrieved from the Dawn JWST Archive (DJA; \citealt{Valentino2023})\footnote{\url{https://dawn-cph.github.io/dja/}}.

\subsection{Photometry}

We measured the NIRCam photometry for COSMOS-3D following the COSMOS2025 framework \citep{Shuntov2025}. Sources were identified in multi-band positive-$\chi^2$ detection images using combined ``hot'' and ``cold'' Source Extraction and Photometry (\texttt{SEP}) detections \citep{Barbary2016}. We measured compact and standard Kron apertures with \texttt{SEP} and \texttt{photutils} \citep{bradley2024} and derived aperture corrections from these measurements. The published COSMOS2025/COSMOS-Web photometric catalog \citep{Shuntov2025} supplies the HST Advanced Camera for Surveys (ACS) F814W and MIRI F770W measurements, as well as the photometric redshifts and their probability density functions (PDFs) used in Section~\ref{sec:photoz_map}. These photometric measurements are the inputs to the component-by-component spectral energy distribution (SED) fits described in Section~\ref{sec:sed}. Champagne et al. (in prep.) will describe the C3D photometry in full.

Two companion systems, C19-5 and C19-8, each contain two resolved emission components (Figure~\ref{fig:c19_c21}(b)). The catalog blends C19-5 into source \texttt{81278}; we measured its components, C19-5a/b (\texttt{81278a/b}), using two-component \texttt{SourceXtractor++} models \citep{Bertin2020,Kummel2020,Kummel2022}, as described in Appendix~\ref{sepp_81278}. The components of C19-8, C19-8a/b (\texttt{257853/257854}), are separately detected, so we use their catalog photometry.

\subsection{\ha~Emission Line Search}
\label{sec:ha_search}

We used the line-search pipeline of \citet{Zou2026}. The parent sample contains COSMOS-Web sources with $m_{\mathrm{F444W}}<28$ whose 16th--84th percentile photometric-redshift intervals overlap $z\simeq4.8$--6.9. We subtracted a local running-median continuum from each one-dimensional spectrum, searched for emission features, and fitted Gaussian templates to measure line fluxes and full widths at half maximum (FWHM). We used \ha~as the primary line, with \nii~and \sii~supporting the identification where covered.

The search required a one-dimensional signal-to-noise ratio of $\mathrm{S/N}_{\mathrm{1D}}>5$ in \ha~and a corresponding $\mathrm{S/N}_{\mathrm{2D}}>3$ detection in the continuum-subtracted two-dimensional spectrum. One-dimensional uncertainties were propagated through reduction and extraction; two-dimensional pixel uncertainties were obtained from the inverse-variance (IVAR) maps as $\sigma_i=\mathrm{IVAR}_i^{-1/2}$. Three assessors inspected both spectra independently. We retained likely or secure emitters with valid one-dimensional fits.

\subsection{Properties of C19 and C21 from Previous Studies}
\label{sec:morph}

The reported stellar masses of C19 and C21 are $\log(M_\star/M_\odot)\simeq9.51$ and 10.11, respectively \citep{Mitsuhashi2024,HerreraCamus2025}. We adopt their ALMA Band~7 continuum flux densities of $0.09\pm0.02$ and $0.41\pm0.15$~mJy, respectively, from \citet{HerreraCamus2025}. Both galaxies appear to lie within a larger structure spanning approximately 25~comoving Mpc (cMpc) \citep{Zou2026}.

C19 was classified as dispersion dominated in the lower-resolution ALPINE \ctwo~data \citep{LeFevre2020} and as a single system in HST/F160W imaging \citep{Mitsuhashi2024}. CRISTAL resolves a coherent \ctwo~velocity gradient aligned with the stellar major axis \citep{HerreraCamus2025,Lee2025,Gomez2026}. Deeper NIRCam imaging reveals a faint secondary component within a disk-like envelope (Figure~\ref{fig:c19_c21}(a)), while the \ctwo~moment-0 map has a faint southern extension \citep{HerreraCamus2025}. Mild azimuthal variations in the gas-phase metallicity gradient have also been reported \citep{Lee2026}.

C21 has \ctwo~emission extending two to three times farther than its rest-frame ultraviolet (UV) emission and disturbed velocity and dispersion maps \citep{Lambert2023}. CRISTAL classifies it as a non-disk system whose gas, dust, and kinematics suggest an interaction or late-stage merger \citep{HerreraCamus2025,Lee2025}.

\section{Results}
\label{sec:results}

\subsection{\ha-Selected Companions around C19 and C21}
\label{sec:member_haes}

\begin{figure*}[htbp]
    \centering
    \includegraphics[width=0.85\textwidth]{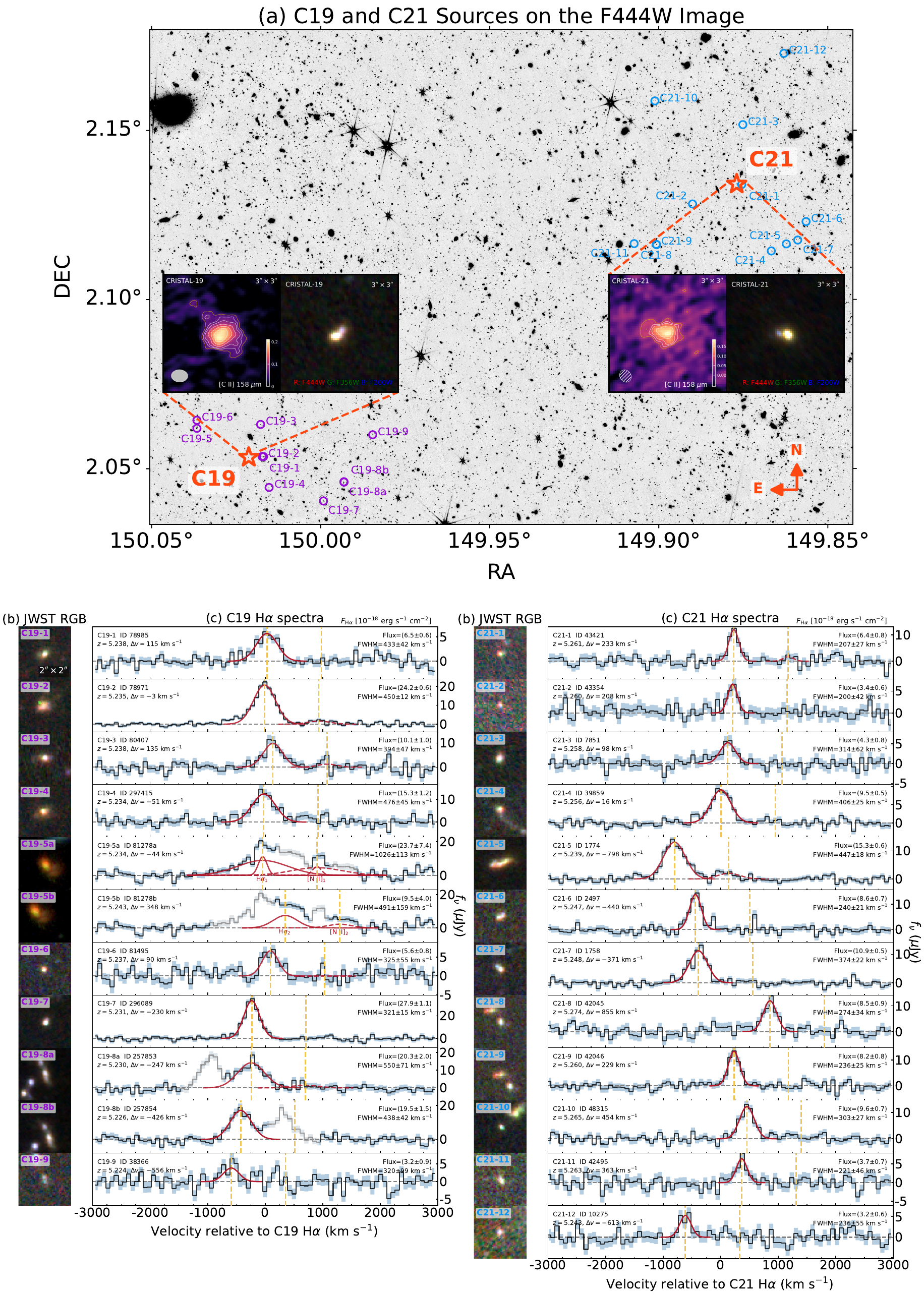}
    \caption{\ha-selected companions of C19 and C21. (a) DJA F444W mosaic \citep{Valentino2023}; red stars mark the targets, and purple/blue symbols mark C19/C21 companions. Insets show ALMA \ctwo~moment-0 maps and target red--green--blue (RGB) images (F444W/F356W/F200W). (b) $2''\times2''$ companion RGB cutouts using the same bands. (c) Grism spectra (black), uncertainties (blue), and Gaussian fits (red), with velocities relative to each target's \ha~redshift. Gray steps and shading mark masked spectral regions and their uncertainties; yellow dashed lines show \ha~and \nii~at each source redshift.}
    \label{fig:c19_c21}
\end{figure*}

We selected companions within 1~physical Mpc (pMpc) in projection and $|\Delta v|<1000~\mathrm{km~s^{-1}}$ of C19 or C21 \citep{Chiang2013,Overzier2016,Shattow2013,Muldrew2015,Muldrew2018,Contini2016}, excluding the targets themselves. This gives 9 systems around C19 and 12 around C21 (Figure~\ref{fig:c19_c21}). C19-5 and C19-8 each contain two resolved \ha-emitting components, giving 11 individually measured components around C19 and 12 around C21. We report the components separately in the spectra, photometry, and SED fits.

\subsection{Overdensity Analysis}
\label{sec:overdensity_analysis}

We identify spectroscopic groups with a friends-of-friends (FoF) algorithm and map projected overdensities using \ha~emitters and COSMOS2025 photometric redshifts \citep{Shuntov2025}. The spectroscopic map uses fixed Gaussian smoothing; the photometric map uses a weighted adaptive kernel \citep{Darvish2015,Brinch2023}.

\begin{figure*}[htbp]
    \centering
    \includegraphics[width=0.8\textwidth]{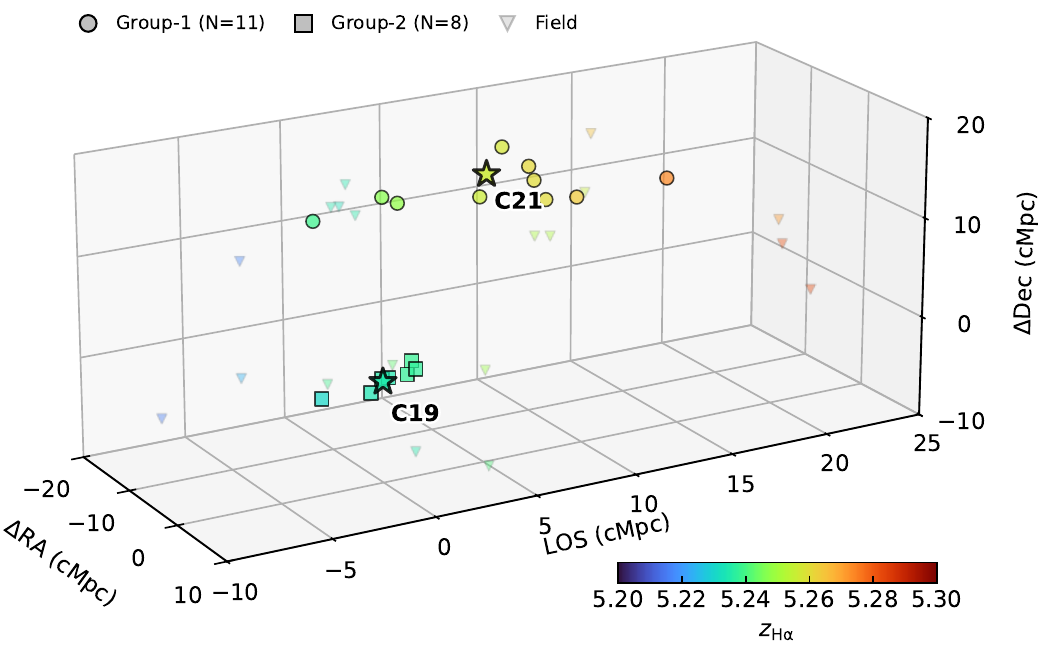}
    \caption{Redshift-space distribution of \ha~emitters at $5.20 < z < 5.30$, with distances in comoving Mpc relative to C19. Circles denote members of the C21 FoF group ($N=11$), and squares denote members of the C19 FoF group ($N=8$). Triangles indicate the remaining \ha~emitters.}
    \label{fig:fof_method}
\end{figure*}

\begin{figure*}[htbp]
    \centering
    \includegraphics[width=\textwidth]{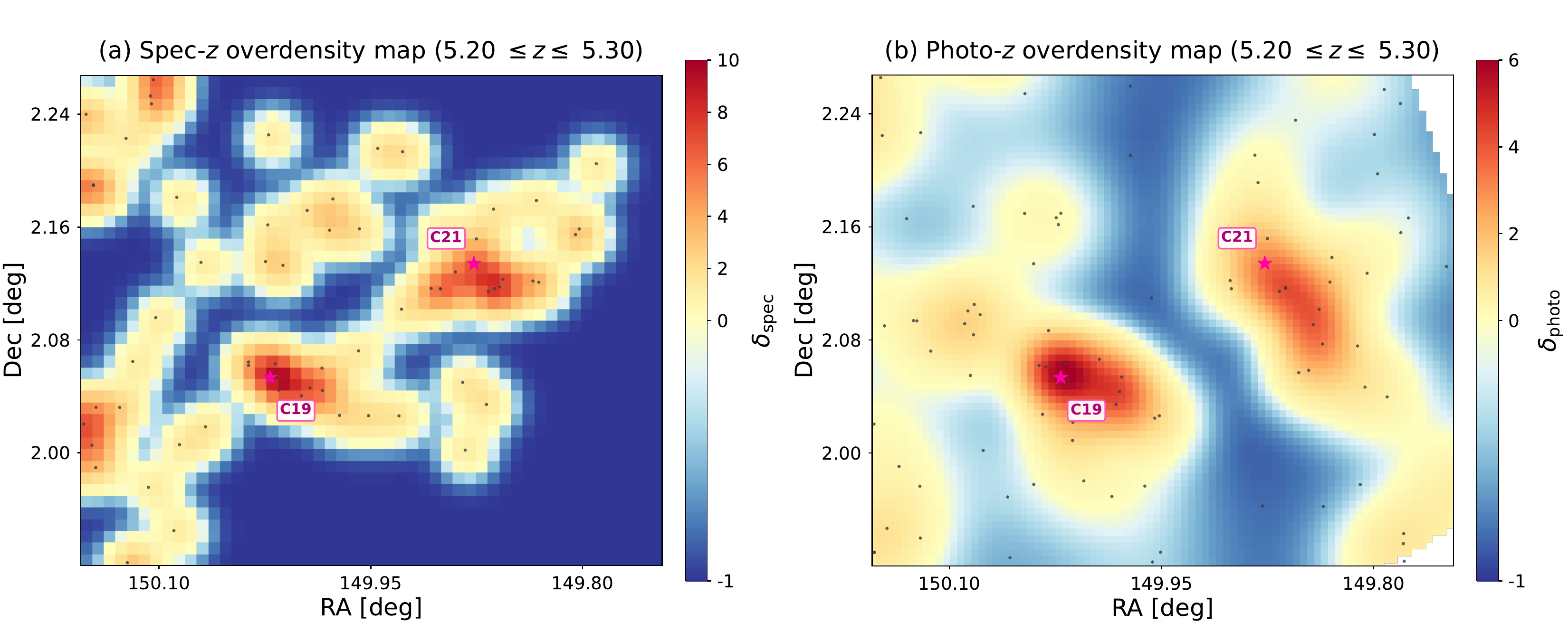}
    \caption{Projected density map at $5.20 \leq z \leq 5.30$: (a) the C3D \ha~spectroscopic sample, (b) the COSMOS2025 photometric sample \citep{Shuntov2025}, using the weighted adaptive kernel method \citep{Brinch2023}.}
    \label{fig:overdensity_map}
\end{figure*}

\subsubsection{Spectroscopic Group Membership}
\label{sec:fof_group}

We apply FoF to \ha~emitters at $5.20\leq z\leq5.30$ across the C3D footprint. Galaxies are linked when their projected separation is below 500~physical kpc and their line-of-sight velocity offset is below $500~\mathrm{km~s^{-1}}$ ($\Delta z\simeq0.01$ at $z\simeq5.2$; \citealt{Sun2024,Helton2024}). Directly or transitively linked galaxies form a group; we retain groups with at least seven members. C19 and C21 belong to separate groups containing 8 and 11 members, respectively, including the targets (Figure~\ref{fig:fof_method}). These groups are separated by approximately 25~comoving Mpc. FoF membership differs from the fixed-window companion selection in Section~\ref{sec:member_haes}.

\subsubsection{Spectroscopic Overdensity Map}
\label{sec:specz_map}

We map \ha~emitters at $5.20\leq z\leq5.30$ on a grid with $0.05$~arcmin pixels, giving every catalog source equal weight and retaining both targets. We smooth with a two-dimensional Gaussian of dispersion $\sigma=0.75$~arcmin, or approximately $0.28$~pMpc at $z=5.25$. The projected overdensity is $\delta_{\mathrm{spec}}=\Sigma_{\mathrm{spec}}/\langle\Sigma_{\mathrm{spec}}\rangle-1$, where $\Sigma_{\mathrm{spec}}$ is the smoothed surface density and $\langle\Sigma_{\mathrm{spec}}\rangle=0.089~\mathrm{arcmin}^{-2}$ is its mean over the adopted normalization region. This is a local density contrast on the smoothing scale, not an aperture-integrated overdensity. Poisson bootstrap realizations assign each source an independent multiplicity with mean unity and recompute the map and its mean with the kernel and mask fixed. We quote the 16th--84th percentiles; these exclude cosmic variance and selection-function uncertainties. At C19 and C21, $\delta_{\mathrm{spec}}=10^{+4}_{-4}$ and $7^{+3}_{-3}$, respectively (Figure~\ref{fig:overdensity_map}(a)).

For detected emitters at $5.20\leq z<5.30$, the median formal $5\sigma$ flux limit is $4.0\times10^{-18}~\mathrm{erg~s^{-1}~cm^{-2}}$, with a 16th--84th percentile range of $(3.0$--$5.9)\times10^{-18}~\mathrm{erg~s^{-1}~cm^{-2}}$, estimated from five times the fitted flux uncertainty. This describes depth at detected-source positions, not catalog completeness. Expanding to $5.0\leq z<5.5$ increases the sample from 151 to 654 emitters and gives a mean comoving number density about 13\% lower for the same effective footprint. The two FoF groups remain unchanged. Neither test accounts for spatial or wavelength-dependent incompleteness.

\subsubsection{Photometric Overdensity and Slice-Width Tests}
\label{sec:photoz_map}

We select COSMOS2025 galaxies \citep{Shuntov2025} whose median photometric redshift, $z_{\mathrm{PDF,med}}$, and peak redshift, $z_{\mathrm{peak}}$, both fall within $5.20\leq z\leq5.30$ and satisfy $|z_{\mathrm{peak}}-z_{\mathrm{PDF,med}}|/|z_{\mathrm{PDF,med}}|\leq0.05$. Each source is weighted by its integrated redshift probability in the slice. The weighted adaptive kernel method \citep{Darvish2015,Brinch2023} represents each source by a Gaussian, with bandwidth optimized by leave-one-out likelihood cross-validation and adapted to local density. We define $\delta_{\mathrm{photo}}=\Sigma_{\mathrm{photo}}/\langle\Sigma_{\mathrm{photo}}\rangle-1$, using the mean over the map footprint. Poisson bootstrap uncertainties are estimated as above, holding the redshift-probability weights and bandwidths fixed. The maps give $\delta_{\mathrm{photo}}=6^{+3}_{-3}$ at C19 and $3^{+2}_{-2}$ at C21 (Figure~\ref{fig:overdensity_map}(b)). Both density estimators have higher point estimates at C19, although their uncertainty intervals overlap.

We recomputed both maps in five slices centered at $z=5.25$, with widths $\Delta z=0.10$--0.50 in steps of 0.10. Both positions remain overdense, with higher point estimates at C19. Across these slices, the mean spectroscopic surface density increases by about a factor of four, more than the local densities. At $\Delta z=0.50$, $\delta_{\mathrm{spec}}$ falls to $3^{+1}_{-1}$ at C19 and $1^{+1}_{-1}$ at C21. The photometric overdensity ranges from $4^{+2}_{-2}$ to $6^{+3}_{-3}$ at C19 and decreases from $3^{+2}_{-2}$ to $1^{+1}_{-1}$ at C21. We use ``nodes'' to denote these distinct galaxy concentrations, without formally identifying cosmic-web nodes.

\subsection{SED Fitting}
\label{sec:sed}

We fitted each companion component at its \ha~spectroscopic redshift using the available HST/ACS F814W, six NIRCam bands (F115W, F150W, F200W, F277W, F356W, F444W), and MIRI F770W photometry. To test the sensitivity of the C19--C21 comparison to star formation history (SFH), we used \texttt{CIGALE} (Code Investigating GALaxy Emission; \citealt{Boquien2019}) and \texttt{BAGPIPES} (Bayesian Analysis of Galaxies for Physical Inference and Parameter EStimation; \citealt{Carnall2018}), each with three SFHs following \citet{Xiao2024}: constant, delayed [$\mathrm{SFR}(t)\propto t\exp(-t/\tau)$], and delayed plus a recent exponentially declining burst. These give six fitting configurations. We compare stellar masses and $\mathrm{SFR}_{100}$, the SED-derived SFR averaged over the preceding 100~Myr. Appendix~\ref{app:sed_settings} lists the age, timescale, burst, and dust parameter ranges.

Both codes use \citet{Bruzual2003} stellar populations and nebular emission, with stellar and gas metallicities varied independently over $Z=0.004$--0.02. \texttt{CIGALE} adopts a Chabrier initial mass function (IMF) \citep{Chabrier2003}, a modified Calzetti law for stellar attenuation \citep{Calzetti2000}, and a Small Magellanic Cloud (SMC) curve for nebular attenuation \citep{Pei1992}, allowing $E(B-V)_{\mathrm{stellar}}/E(B-V)_{\mathrm{line}}=0.5$ or 1.0. \texttt{BAGPIPES} adopts a Kroupa IMF \citep{Kroupa2002} and the \citet{Salim2018} attenuation prescription. The non-SFH parameter settings are unchanged between the three runs within each code. Comparisons between codes therefore also include differences in IMF, attenuation, and priors.

We use the \texttt{CIGALE} constant-SFH results for our analysis (Table~\ref{tab:sed_parameter_comparison}). For the three \texttt{CIGALE} runs, we calculate the Bayesian information criterion (BIC), $\mathrm{BIC}=\chi^2_{\mathrm{min}}+k\ln N$, where $k$ is the number of fitted parameters and $N$ the number of photometric bands. The constant SFH gives the lowest BIC for 20 of 23 components; the remaining three prefer the delayed SFH by only 0.5--1.4. The median per-source range in stellar mass across SFHs is 0.13~dex for \texttt{CIGALE} and 0.32~dex for \texttt{BAGPIPES}. All six configurations give higher median companion masses and SFRs around C19, although the contrast varies. Table~\ref{tab:sed_robustness} summarizes these differences. We use these component masses for the halo estimates and the corresponding nebular color excesses for \ha~attenuation corrections. The individual fits are shown in Figures~\ref{fig:sed_fits}--\ref{fig:sed_fits_06} (Appendix~\ref{app:sed_fits}).

\begin{table*}[t]
\centering
\caption{H$\alpha$ spectral measurements and CIGALE constant-SFH results.}
\label{tab:sed_parameter_comparison}
\scriptsize
\setlength{\tabcolsep}{4pt}
\begin{tabular}{lcccccccc}
\hline
ID & $z_{\mathrm{H}\alpha}$ & $f_{\mathrm{H}\alpha}$ & $E(B-V)_{\rm line}$ & $\mathrm{SFR}_{\mathrm{H}\alpha,\rm corr}$ & FWHM & $\log(M_\star/M_\odot)$ & $\mathrm{SFR}_{100}$ & Age \\
 & & $(10^{-18}~\mathrm{erg~s^{-1}~cm^{-2}})$ & (mag) & $(M_\odot~\mathrm{yr}^{-1})$ & $(\mathrm{km~s}^{-1})$ & & $(M_\odot~\mathrm{yr}^{-1})$ & (Myr) \\
\hline
\multicolumn{9}{c}{CRISTAL-19} \\
78985 & 5.238 & $6.5^{+0.6}_{-0.6}$ & $0.21^{+0.13}_{-0.13}$ & $14^{+4}_{-4}$ & $433^{+42}_{-42}$ & $9.67^{+0.19}_{-0.19}$ & $27^{+20}_{-20}$ & $404^{+327}_{-327}$ \\
78971 & 5.235 & $24.2^{+0.6}_{-0.6}$ & $0.24^{+0.12}_{-0.12}$ & $57^{+14}_{-14}$ & $450^{+12}_{-12}$ & $9.62^{+0.29}_{-0.29}$ & $98^{+55}_{-55}$ & $129^{+210}_{-210}$ \\
80407 & 5.238 & $10.1^{+1.0}_{-1.0}$ & $0.17^{+0.10}_{-0.10}$ & $21^{+5}_{-5}$ & $394^{+47}_{-47}$ & $8.96^{+0.34}_{-0.34}$ & $35^{+17}_{-17}$ & $67^{+124}_{-124}$ \\
297415 & 5.234 & $15.3^{+1.2}_{-1.2}$ & $0.15^{+0.12}_{-0.12}$ & $30^{+8}_{-8}$ & $476^{+45}_{-45}$ & $9.61^{+0.17}_{-0.17}$ & $20^{+21}_{-21}$ & $512^{+341}_{-341}$ \\
81278a & 5.234 & $23.7^{+7.4}_{-7.4}$ & $0.49^{+0.15}_{-0.15}$ & $93^{+41}_{-41}$ & $1026^{+113}_{-113}$ & $10.47^{+0.24}_{-0.24}$ & $163^{+96}_{-96}$ & $390^{+312}_{-312}$ \\
81278b & 5.243 & $9.5^{+4.0}_{-4.0}$ & $0.39^{+0.05}_{-0.05}$ & $31^{+13}_{-13}$ & $491^{+159}_{-159}$ & $9.22^{+0.31}_{-0.31}$ & $110^{+31}_{-31}$ & $23^{+45}_{-45}$ \\
81495 & 5.237 & $5.6^{+0.8}_{-0.8}$ & $0.22^{+0.10}_{-0.10}$ & $13^{+3}_{-3}$ & $325^{+55}_{-55}$ & $8.78^{+0.34}_{-0.34}$ & $25^{+11}_{-11}$ & $75^{+152}_{-152}$ \\
296089 & 5.231 & $27.9^{+1.1}_{-1.1}$ & $0.08^{+0.06}_{-0.06}$ & $48^{+6}_{-6}$ & $321^{+15}_{-15}$ & $8.95^{+0.23}_{-0.23}$ & $27^{+12}_{-12}$ & $60^{+64}_{-64}$ \\
257853 & 5.230 & $20.3^{+2.0}_{-2.0}$ & $0.09^{+0.08}_{-0.08}$ & $35^{+6}_{-6}$ & $550^{+71}_{-71}$ & $9.91^{+0.19}_{-0.19}$ & $71^{+37}_{-37}$ & $212^{+165}_{-165}$ \\
257854 & 5.226 & $19.5^{+1.5}_{-1.5}$ & $0.09^{+0.07}_{-0.07}$ & $34^{+5}_{-5}$ & $438^{+42}_{-42}$ & $9.43^{+0.23}_{-0.23}$ & $51^{+24}_{-24}$ & $99^{+95}_{-95}$ \\
38366 & 5.224 & $3.2^{+0.9}_{-0.9}$ & $0.07^{+0.06}_{-0.06}$ & $5^{+2}_{-2}$ & $320^{+99}_{-99}$ & $8.56^{+0.28}_{-0.28}$ & $7^{+3}_{-3}$ & $109^{+122}_{-122}$ \\
\hline
\multicolumn{9}{c}{CRISTAL-21} \\
43421 & 5.261 & $6.4^{+0.8}_{-0.8}$ & $0.34^{+0.14}_{-0.14}$ & $19^{+6}_{-6}$ & $207^{+27}_{-27}$ & $8.81^{+0.31}_{-0.31}$ & $21^{+10}_{-10}$ & $67^{+118}_{-118}$ \\
43354 & 5.260 & $3.4^{+0.6}_{-0.6}$ & $0.07^{+0.06}_{-0.06}$ & $6^{+1}_{-1}$ & $200^{+42}_{-42}$ & $7.87^{+0.23}_{-0.23}$ & $4^{+2}_{-2}$ & $23^{+25}_{-25}$ \\
7851 & 5.258 & $4.3^{+0.8}_{-0.8}$ & $0.10^{+0.09}_{-0.09}$ & $8^{+2}_{-2}$ & $314^{+62}_{-62}$ & $9.15^{+0.23}_{-0.23}$ & $15^{+9}_{-9}$ & $201^{+188}_{-188}$ \\
39859 & 5.256 & $9.5^{+0.5}_{-0.5}$ & $0.08^{+0.09}_{-0.09}$ & $16^{+3}_{-3}$ & $406^{+25}_{-25}$ & $9.50^{+0.20}_{-0.20}$ & $16^{+12}_{-12}$ & $444^{+319}_{-319}$ \\
1774 & 5.239 & $15.3^{+0.6}_{-0.6}$ & $0.58^{+0.06}_{-0.06}$ & $72^{+9}_{-9}$ & $447^{+18}_{-18}$ & $9.52^{+0.22}_{-0.22}$ & $233^{+55}_{-55}$ & $22^{+33}_{-33}$ \\
2497 & 5.247 & $8.6^{+0.7}_{-0.7}$ & $0.13^{+0.08}_{-0.08}$ & $16^{+3}_{-3}$ & $240^{+21}_{-21}$ & $8.59^{+0.38}_{-0.38}$ & $16^{+6}_{-6}$ & $53^{+103}_{-103}$ \\
1758 & 5.248 & $10.9^{+0.5}_{-0.5}$ & $0.34^{+0.08}_{-0.08}$ & $32^{+5}_{-5}$ & $374^{+22}_{-22}$ & $8.58^{+0.12}_{-0.12}$ & $36^{+6}_{-6}$ & $13^{+10}_{-10}$ \\
42045 & 5.274 & $8.5^{+0.9}_{-0.9}$ & $0.10^{+0.09}_{-0.09}$ & $15^{+3}_{-3}$ & $274^{+34}_{-34}$ & $8.84^{+0.26}_{-0.26}$ & $7^{+5}_{-5}$ & $218^{+224}_{-224}$ \\
42046 & 5.260 & $8.2^{+0.8}_{-0.8}$ & $0.38^{+0.13}_{-0.13}$ & $26^{+7}_{-7}$ & $236^{+25}_{-25}$ & $9.26^{+0.37}_{-0.37}$ & $40^{+25}_{-25}$ & $150^{+241}_{-241}$ \\
48315 & 5.265 & $9.6^{+0.7}_{-0.7}$ & $0.16^{+0.06}_{-0.06}$ & $20^{+3}_{-3}$ & $303^{+27}_{-27}$ & $8.17^{+0.16}_{-0.16}$ & $13^{+2}_{-2}$ & $13^{+13}_{-13}$ \\
42495 & 5.263 & $3.7^{+0.7}_{-0.7}$ & $0.09^{+0.09}_{-0.09}$ & $6^{+2}_{-2}$ & $221^{+46}_{-46}$ & $8.83^{+0.24}_{-0.24}$ & $5^{+4}_{-4}$ & $305^{+273}_{-273}$ \\
10275 & 5.243 & $3.2^{+0.6}_{-0.6}$ & $0.14^{+0.08}_{-0.08}$ & $6^{+2}_{-2}$ & $236^{+55}_{-55}$ & $8.39^{+0.25}_{-0.25}$ & $6^{+3}_{-3}$ & $74^{+94}_{-94}$ \\
\hline
\multicolumn{9}{@{}l@{}}{Within each group, sources follow the top-to-bottom order in Figure~\ref{fig:c19_c21}(c).} \\
\multicolumn{9}{@{}l@{}}{Stellar parameters for \texttt{81278b} depend on whether its extended emission traces stars or an ionized nebula; see Section~\ref{sec:discussion}.} \\
\end{tabular}
\end{table*}

\subsection{Dust-Corrected \ha~SFRs}
\label{sec:sfr_corr}

We correct each \ha~flux using the nebular color excess from the constant-SFH fits: $A_{\mathrm{H\alpha}}=2.20E(B-V)_{\mathrm{line}}$, $F_{\mathrm{H\alpha,corr}}=F_{\mathrm{H\alpha,obs}}10^{0.4A_{\mathrm{H\alpha}}}$, and $L_{\mathrm{H\alpha,corr}}=4\pi D_L^2F_{\mathrm{H\alpha,corr}}$. The \citet{Kennicutt1998} calibration, rescaled to a Chabrier IMF \citep{Chabrier2003,Madau2014}, gives $\mathrm{SFR}_{\mathrm{H\alpha,corr}}=5.0\times10^{-42}L_{\mathrm{H\alpha,corr}}$, with luminosity in $\mathrm{erg~s^{-1}}$ and SFR in $M_\odot~\mathrm{yr}^{-1}$. The companion medians are $31^{+6}_{-9}$ and $16^{+3}_{-5}~M_\odot~\mathrm{yr}^{-1}$ around C19 and C21, respectively (Table~\ref{tab:sed_parameter_comparison}); errors are the 16th--84th percentiles of bootstrap medians.

\section{Discussion}
\label{sec:discussion}

\subsection{Contrasting Assembly in the C19 and C21 Nodes}

\begin{figure}[htbp]
    \centering
    \includegraphics[width=\columnwidth]{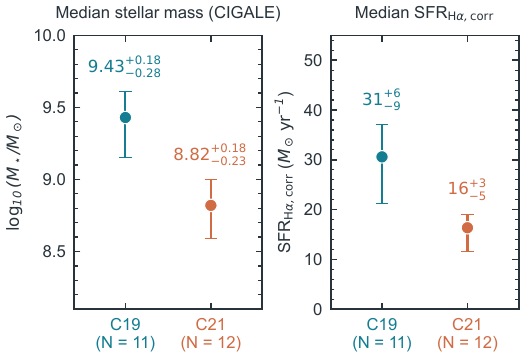}
    \caption{Median companion stellar masses (left) and dust-corrected \ha~SFRs (right). Masses and nebular color excesses come from the \texttt{CIGALE} constant-SFH fits. Errors are the 16th--84th percentiles of bootstrap medians.}
    \label{fig:median_mass_sfr}
\end{figure}

Overdense environments provide opportunities for galaxy interactions and mergers, but a rich environment need not coincide with the most massive central galaxy. C19 and C21 illustrate this distinction: both occupy overdense nodes, yet C19 has the more massive, actively star-forming companions, while C21 itself is more massive. The component-resolved constant-SFH fits give median companion masses of $\log(M_\star/M_\odot)=9.43^{+0.18}_{-0.28}$ and $8.82^{+0.18}_{-0.23}$, respectively. This ordering persists across all six SED configurations (Section~\ref{sec:sed}). Figure~\ref{fig:median_mass_sfr} compares their median companion masses and dust-corrected \ha~SFRs.

To compare the companions with field galaxies, we use stellar masses and 100~Myr-averaged SFRs from the same broadband SED-fitting catalog for both populations. We compared companions within 1~pMpc and $|\Delta v|<1000~\mathrm{km~s^{-1}}$ with control HAEs beyond 2~pMpc of both nodes and outside rich FoF groups. Usable estimates are available for eight cataloged C19 companions and twelve C21 companions. Their median masses are $\log(M_\star/M_\odot)=9.76^{+0.11}_{-0.24}$ and $9.11^{+0.17}_{-0.08}$, compared with $9.33^{+0.04}_{-0.11}$ for 83 controls at $5.2\leq z<5.3$ and $9.17^{+0.11}_{-0.09}$ for 44 controls at $5.4\leq z<5.5$. Matching controls within 0.3~dex in stellar mass gives median $\mathrm{SFR}_{100}$ offsets from the same-redshift controls of $0.13^{+0.08}_{-0.08}$ and $0.01^{+0.02}_{-0.10}$~dex for C19 and C21. These modest offsets suggest that C19's higher absolute SFRs largely reflect its more massive companion population, rather than a strong enhancement at fixed mass. The uncertainties include bootstrap resampling of companions and controls, but not SED-model systematics. This mass-matched comparison uses SED-derived SFRs averaged over 100~Myr. Figure~\ref{fig:median_mass_sfr} instead shows companion masses from our component-by-component \texttt{CIGALE} fits and SFRs derived from dust-corrected \ha~luminosities.

The COSMOS2025 photometric-redshift sample \citep{Shuntov2025} provides a complementary comparison. Applying the criteria in Section~\ref{sec:photoz_map} over $5.10\leq z\leq5.40$ yields 851 sources, including 18 neighbors within 1~pMpc of C19 and 14 around C21, excluding the targets. Their PDF-weighted median masses are $\log(M_\star/M_\odot)=9.57^{+0.09}_{-0.25}$ and $9.35^{+0.06}_{-0.33}$, and their median $\mathrm{SFR}_{100}$ values are $9.89^{+5.17}_{-4.54}$ and $7.30^{+6.17}_{-2.35}\,M_\odot\,\mathrm{yr}^{-1}$. The 257 field galaxies with $\delta_{\mathrm{photo}}\leq0$, excluding both neighbor samples, have corresponding medians of $9.24^{+0.03}_{-0.05}$ and $4.41^{+0.41}_{-0.29}\,M_\odot\,\mathrm{yr}^{-1}$. Both neighbor medians are again higher around C19, although their uncertainty intervals overlap. This comparison supports the ordering without establishing a significant photometric-sample difference between the nodes.

The C19 environment contains two resolved two-component systems, C19-5a/b (\texttt{81278a/b}) and C19-8a/b (\texttt{257853/257854}; Figure~\ref{fig:c19_c21}(b)), whose morphologies suggest possible interactions or mergers. Our two-component photometric modelling and tests of how the derived stellar masses and SFRs depend on the SED-fitting assumptions are described in Appendices~\ref{sepp_81278} and \ref{app:sed_settings}. C19-5a also shows a broad observed \ha~profile, with $\mathrm{FWHM}=1026\pm113~\mathrm{km~s^{-1}}$ (Figure~\ref{fig:c19_c21}(c); Table~\ref{tab:sed_parameter_comparison}), suggesting possible AGN activity within this multicomponent system. A detailed analysis of C19-5 and its candidate AGN-driven outflow is presented by \citet{Wei2026}.

C21 is the more massive target with disturbed \ctwo~morphology and non-disk kinematics, interpreted as an interaction or late-stage merger \citep{Lambert2023,HerreraCamus2025,Lee2025}. These observations are consistent with mergers having contributed to its stellar buildup, while its remaining \ha-selected companions are less massive and less actively star-forming than those around C19.

We quantify the stellar-mass distribution using the same catalog-based SED mass estimates for each target and its selected companions. C19 contributes $4.5^{+0.9}_{-0.7}\%$ of the combined measured mass of the target and its selected companions, compared with $39^{+5}_{-5}\%$ for C21. Even after excluding its most massive companion, \texttt{81278}, C19 contains less than 9\% of the combined measured stellar mass. C19's companions therefore dominate its local stellar-mass budget, whereas C21 itself contains a much larger share, revealing a contrast in mass distribution beyond the difference in companion counts. These fractions describe the selected populations; one C19 companion lacks a matching mass estimate, and undetected members are not included.

The stellar-mass fractions add a quantitative distinction to the environmental trends of \citet{Zou2026}: two overdense nodes at nearly the same epoch differ in how much of their measured stellar mass resides within the target galaxy. Figures~\ref{fig:c19_c21} and \ref{fig:median_mass_sfr} locate the accompanying growth in different parts of these populations. Around C19, substantial mass and ongoing star formation remain in separate companions, including candidate interacting systems; in C21, the larger target mass fraction accompanies evidence for a late-stage merger within the target itself. The modest mass-matched SFR offsets suggest that this contrast primarily concerns accumulated stellar mass, rather than a large enhancement of recent star formation at fixed mass. C21's broad \ha~component is also not sufficient to identify an AGN: follow-up NIRSpec spectra and Baldwin--Phillips--Terlevich (BPT) diagnostics favour intense central star formation as its origin (M. Aravena et al., in prep.).

\subsection{Halo Growth and Possible Cluster Descendants}
\label{sec:halo_mass_est}

We converted individual stellar masses to halo masses using the stellar mass--halo mass (SMHM) relation of \citet{Behroozi2019} and summed the estimates for each target and its selected companions. We obtain $\log(\sum_i M_{\mathrm{halo},i}/M_\odot)=12.79^{+0.05}_{-0.05}$ for C19 and $12.50^{+0.05}_{-0.05}$ for C21. Expressed in halo-mass terms, the observed stellar-mass distribution therefore gives the opposite ranking from the target masses: C19 has the larger summed estimate despite its less massive target. These sums are derived from the same stellar masses and describe the selected members, rather than independently measured masses of single halos. The more massive target is consequently a poor guide to the mass associated with its surrounding population in this pair.

\begin{figure}[htbp]
    \centering
    \includegraphics[width=\columnwidth]{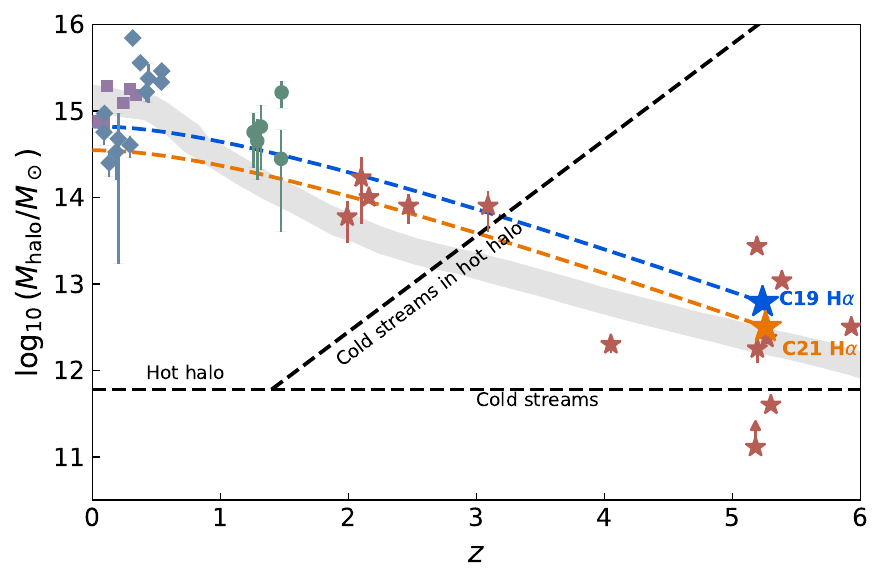}
    \caption{Summed halo-mass estimates for C19 (blue star) and C21 (orange star), with illustrative growth tracks \citep{McBride2009,Fakhouri2010} reaching $\log(M_{\mathrm{halo}}/M_\odot)=14.82$ and 14.55 at $z=0$. The tracks assume that each selected population assembles into a common halo. Red stars show literature protoclusters \citep{Casey2016,Helton2024b}; green circles, blue diamonds, and purple squares show clusters from \citet{Zohren2022}, \citet{Sereno2025}, and \citet{Williams2026}, respectively. The gray band shows Coma-like halo growth \citep{Chiang2013}. Black dashed boundaries separate cold-stream and hot-halo accretion regimes for individual halos \citep{Dekel2006}.}
    \label{fig:halo_evolution}
\end{figure}

Figure~\ref{fig:halo_evolution} connects these local mass budgets to possible later structures. Conditional on each selected population assembling into a common halo, the prescriptions of \citet{McBride2009,Fakhouri2010} give tracks reaching $\log(M_{\mathrm{halo}}/M_\odot)=14.82^{+0.05}_{-0.05}$ and $14.55^{+0.05}_{-0.05}$ at $z=0$. Their overlap with the mass range of the observed $z\sim2$ protocluster systems in the figure illustrates a possible pathway from these early nodes to cluster-scale structures. The distinct groups and density concentrations in Figures~\ref{fig:fof_method} and \ref{fig:overdensity_map} may nevertheless belong to the same larger complex. Simulated protoclusters contain multiple assembling halos and extended star-forming populations through $z\sim2$ \citep{Muldrew2015,chiang17}. If the two nodes share a final cluster, their curves are separate extrapolations of local mass budgets, not predictions of two final clusters. SMHM scatter, incomplete membership, shared host halos, and variations in accretion further limit the descendant estimates.

Hierarchical models of brightest cluster galaxy (BCG) formation distinguish early star formation in separate progenitors from their later assembly through mergers \citep{delucia07}. The mass distributions measured here provide early observational constraints on such histories: C19 has a larger reservoir of stellar mass in companions, while C21 contains a larger share within a merging target. Neither target is identified as the eventual BCG. Instead, the observations provide two different distributions of stellar mass and ongoing growth that models connecting early massive galaxies to later cluster populations must account for. Reproducing the target mass alone would miss the contrasting companion populations and the locations of ongoing assembly.

\section{Summary}
\label{sec:summary}

We investigated the environments of CRISTAL-19 (C19) and CRISTAL-21 (C21) at $z\simeq5.2$ using COSMOS-3D NIRCam F444W slitless spectroscopy, multi-band imaging, and ALMA observations. Our main findings are as follows.

\begin{enumerate}

\item We find that the two CRISTSAL galaxies: C19 and C21, occupy separate overdense nodes, assigned to distinct FoF groups approximately 25~cMpc apart in redshift space. Both spectroscopic and photometric maps show density enhancements: $\delta_{\mathrm{spec}}=10^{+4}_{-4}$ and $7^{+3}_{-3}$, and $\delta_{\mathrm{photo}}=6^{+3}_{-3}$ and $3^{+2}_{-2}$, respectively. 

\item The \ha-selected companions around C19 have higher median stellar masses and SFRs than those around C21. Its environment contains the candidate interacting pair C19-8a/b and the multicomponent system C19-5a/b. The environemnt the C19 have mmore merging morphology in the compoanisoons. C19-5 shows two components and the major one possibly shows the AGN features given the broad component in the \ha~lines. These systems identify sites of ongoing assembly and possible nuclear activity outside C19 itself. The higher absolute companion SFRs largely reflect their greater stellar masses, rather than a strong enhancement at fixed mass.

\item Of the two massive central galaxies, C21 is more massive than C19, with disturbed \ctwo~morphology and kinematics suggesting an interacting or late-stage merger system. Using the same catalog-based SED mass estimates for each central galaxy and its companions, we find that C19 and C21 contribute $4.5^{+0.9}_{-0.7}\%$ and $39^{+5}_{-5}\%$ of the measured stellar mass in their respective populations. This contrast is consistent with more mass having assembled within C21, while substantial growth continues in separate galaxies around C19.

\item We find that the adopted halo estimates and growth histories allow cluster-scale descendants. Summing the member halo masses inferred from the SMHM relation gives $\log(\sum_i M_{\mathrm{halo},i}/M_\odot)=12.79\pm0.05$ for C19 and $12.50\pm0.05$ for C21. The illustrative tracks reach $\log(M_{\mathrm{halo}}/M_\odot)=14.82\pm0.05$ and $14.55\pm0.05$ at $z=0$, respectively, conditional on each selected population assembling into a common halo.

\end{enumerate}

The two nodes have opposite target and companion mass rankings, with only $4.5\%$ of the measured stellar mass in C19 itself compared with $39\%$ in C21. Their modest mass-matched SFR offsets show that the contrast primarily reflects the distribution of accumulated stellar mass. Together with the resolved companions and merger evidence, these measurements locate different sites of ongoing assembly within neighbouring overdense regions, extending the companion-count trends of \citet{Zou2026}.

\begin{acknowledgments}
We thank the COSMOS-3D, COSMOS-Web, and ALMA-CRISTAL collaborations for the data used in this work. SZG and SZ thank Yongda Zhu and Nicolas Laporte for fruitful discussions and Jackie Champagne for providing the photometric catalog.

SZG, SZ, NP and YR acknowledge support from the National Natural Science Foundation of China (grant no.\ 12673018), Chinese Academy of Sciences (no. E529540101), CASSACA Key Research Project E52H540101 and the NAOC Research Program E5TG540101. This work is sponsored by the Chinese Academy of Sciences (CAS), through grant Y329011V01 to the CAS South America Center for Astronomy (CASSACA). MA acknowledges support from FONDECYT grant no.\ 1252054, ANID Basal Project FB210003, ANID MILENIO NCN2024\_112, and ANID + Vinculaci\'on Internacional + FOVI250261. L.L.L. acknowledges financial support from the European Research Council (ERC) Advanced Grant under the European Union's (EU's) Horizon Europe research and innovation programme (grant agreement AdG GALPHYS, No. 101055023). Views and opinions expressed are, however, those of the author(s) only and do not necessarily reflect those of the EU or the ERC. Neither the EU nor the granting authority can be held responsible for them. JM gratefully acknowledges support from ANID MILENIO NCN2024\_112.

This work is based on observations made with the NASA/ESA/CSA James Webb Space Telescope. The data were obtained from the Mikulski Archive for Space Telescopes at the Space Telescope Science Institute, which is operated by the Association of Universities for Research in Astronomy, Inc., under NASA contract NAS 5-03127 for JWST. These observations are associated with programs \#5893 and \#1727. Support for program \#5893 was provided by NASA through a grant from the Space Telescope Science Institute, which is operated by the Association of Universities for Research in Astronomy, Inc., under NASA contract NAS 5-03127. The specific observations analyzed can be accessed via \dataset[doi:10.17909/k5th-zb31]{https://doi.org/10.17909/k5th-zb31}. We acknowledge the strong support provided by the program coordinator Alison Vick and instrument reviewers Brian Brooks and Jonathan Aguilar.

The JWST/NIRCam imaging in Figure~\ref{fig:c19_c21} was retrieved from the Dawn JWST Archive (DJA). DJA is an initiative of the Cosmic Dawn Center (DAWN), which is funded by the Danish National Research Foundation under grant DNRF140.

This paper makes use of the following ALMA data: ADS/JAO.ALMA\#2021.1.00280.L and ADS/JAO.ALMA\#2018.1.01359.S. ALMA is a partnership of ESO (representing its member states), NSF (USA), and NINS (Japan), together with NRC (Canada), NSTC and ASIAA (Taiwan), and KASI (Republic of Korea), in cooperation with the Republic of Chile. The Joint ALMA Observatory is operated by ESO, AUI/NRAO, and NAOJ. The National Radio Astronomy Observatory and Green Bank Observatory are facilities of the U.S. National Science Foundation operated under cooperative agreement by Associated Universities, Inc.
\end{acknowledgments}

\facilities{JWST/NIRCam, ALMA, HST}

\software{\texttt{astropy} \citep{2013A&A...558A..33A,2018AJ....156..123A,2022ApJ...935..167A}, \texttt{Cloudy} \citep{2013RMxAA..49..137F},  \texttt{SourceXtractor++} \citep{Bertin2020,Kummel2020,Kummel2022}, \texttt{CIGALE} \citep{Boquien2019}, \texttt{BAGPIPES} \citep{Carnall2018}, \texttt{grizli} \citep{Brammer2023}, \texttt{SEP} \citep{Barbary2016}, \texttt{photutils} \citep{bradley2024}}

\clearpage
\restartappendixnumbering
\appendix

\section{Two-component Photometry of C19-5}
\label{sepp_81278}

We measured C19-5a/b in six NIRCam bands with S\'ersic models convolved with the point-spread function (PSF) in \texttt{SourceXtractor++} \citep{Bertin2020,Kummel2020,Kummel2022}. We used $51\times51$ pixel cutouts at $0.05$~arcsec pixel$^{-1}$, subtracted local backgrounds with iterative $3\sigma$ clipping, and converted the images to nJy pixel$^{-1}$. Pixels contaminated by a bright neighbour were replaced with Gaussian noise matched to the background root-mean-square (rms) noise. Detection used $D=\sum_b I_b/\sigma_{\mathrm{bg},b}/\sqrt{3}$ over F277W, F356W, and F444W, where $I_b$ and $\sigma_{\mathrm{bg},b}$ are the background-subtracted image and background rms in each band. Independent band noises give approximately unit background variance. We restricted detection to the central region by replacing peripheral pixels with unit Gaussian noise and required five connected pixels above three times the normalized noise.

\begin{figure*}[htbp]
    \centering
    \includegraphics[width=0.75\textwidth]{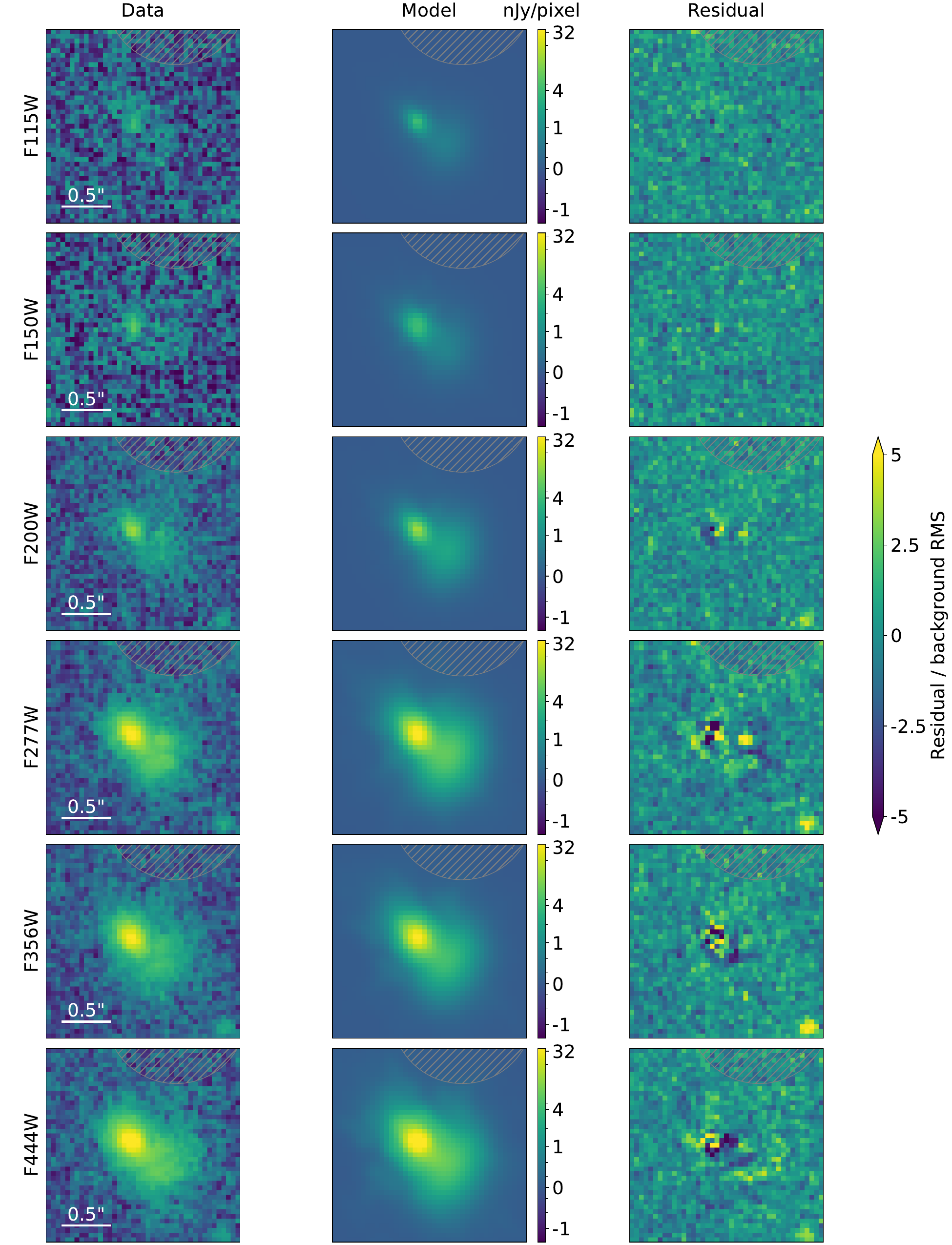}
    \caption{Observed images, two-component \texttt{SourceXtractor++} models, and residuals for C19-5.}
    \label{fig:sepp}
\end{figure*}

Empirical PSFs were constructed from four stars from Gaia Data Release~3 (DR3) within 3~arcmin \citep{Lizj2026}. We masked neighbours in $121\times121$ pixel stellar cutouts, subtracted local backgrounds, fitted Gaussian centroids, and resampled by a factor of three. Stars were normalized within $2\arcsec$ apertures and median-stacked; the final PSFs have unit integral. We jointly fitted F277W, F356W, and F444W with two S\'ersic components, sharing each component's structural parameters between bands and fitting its fluxes independently. We then fixed the structures to measure all six bands. Figure~\ref{fig:sepp} shows the models and residuals. Formal errors exclude uncertainty in the fixed structural parameters and PSF modelling.

\section{SED-fitting Parameters and Robustness}
\label{app:sed_settings}

For the SFH configurations described in Section~\ref{sec:sed}, the main-component age ranged from 10 to 1000 Myr, with $\tau=10$--3000 Myr for the delayed SFHs. The recent burst had a fixed e-folding timescale of 10 Myr. In \texttt{CIGALE}, its age was 1, 3, or 5 Myr, and its mass fraction ranged from 0 to 0.5 of the total formed stellar mass. In \texttt{BAGPIPES}, its age had a uniform prior over 1--5 Myr; it shared the main component's stellar metallicity, while its formed mass was fitted independently.

For \texttt{CIGALE}, stellar and gas metallicities were independently sampled at $Z=0.004$, 0.008, and 0.02, with $\log U$ ranging from -3.5 to -1.0. Stellar and nebular emission were combined with dust attenuation and re-emission under an energy-balance prescription. The attenuation-law slope modification ranged from -0.6 to 0.2, with no UV bump, and $E(B-V)_{\mathrm{line}}=0$--1.0.

For \texttt{BAGPIPES}, stellar and gas metallicities were fitted independently over $0.2\leq Z/Z_\odot\leq1.0$, with $Z_\odot=0.02$. Nebular line and continuum emission were included using pre-computed \texttt{Cloudy} grids with $-3.5\leq\log U\leq-1.0$ \citep{Gunasekera2025}. The attenuation parameters were $A_V=0$--4 mag and a slope parameter from -0.6 to 0.2, with no UV bump.

\begin{table*}[t]
\centering
\caption{Companion-population contrasts across SED-fitting configurations.}
\label{tab:sed_robustness}
\scriptsize
\setlength{\tabcolsep}{8pt}
\begin{tabular}{llcc}
\hline
Code & SFH & $\Delta\mathrm{median}[\log(M_\star/M_\odot)]$ & $\Delta\log[\mathrm{median}(\mathrm{SFR}_{100})]$ \\
 & & (dex) & (dex) \\
\hline
\texttt{CIGALE} & Constant & $0.61$ & $0.35$ \\
\texttt{CIGALE} & Delayed & $0.69$ & $0.37$ \\
\texttt{CIGALE} & Delayed + burst & $0.77$ & $0.54$ \\
\texttt{BAGPIPES} & Constant & $0.60$ & $0.46$ \\
\texttt{BAGPIPES} & Delayed & $0.60$ & $0.56$ \\
\texttt{BAGPIPES} & Delayed + burst & $1.12$ & $1.08$ \\
\hline
\multicolumn{4}{@{}l@{}}{Offsets are C19 minus C21; all six configurations give higher median masses and SFRs around C19.} \\
\multicolumn{4}{@{}l@{}}{Contrasts use the rounded component estimates from each SED-fitting configuration.} \\
\multicolumn{4}{@{}l@{}}{Individual posterior uncertainties are not propagated. $\mathrm{SFR}_{100}$ is SED-derived, not the \ha~SFR.} \\
\end{tabular}
\end{table*}

\section{Individual SED Fits}
\label{app:sed_fits}

\begin{figure*}[p]
    \centering
    \includegraphics[width=\textwidth,height=0.88\textheight,keepaspectratio]{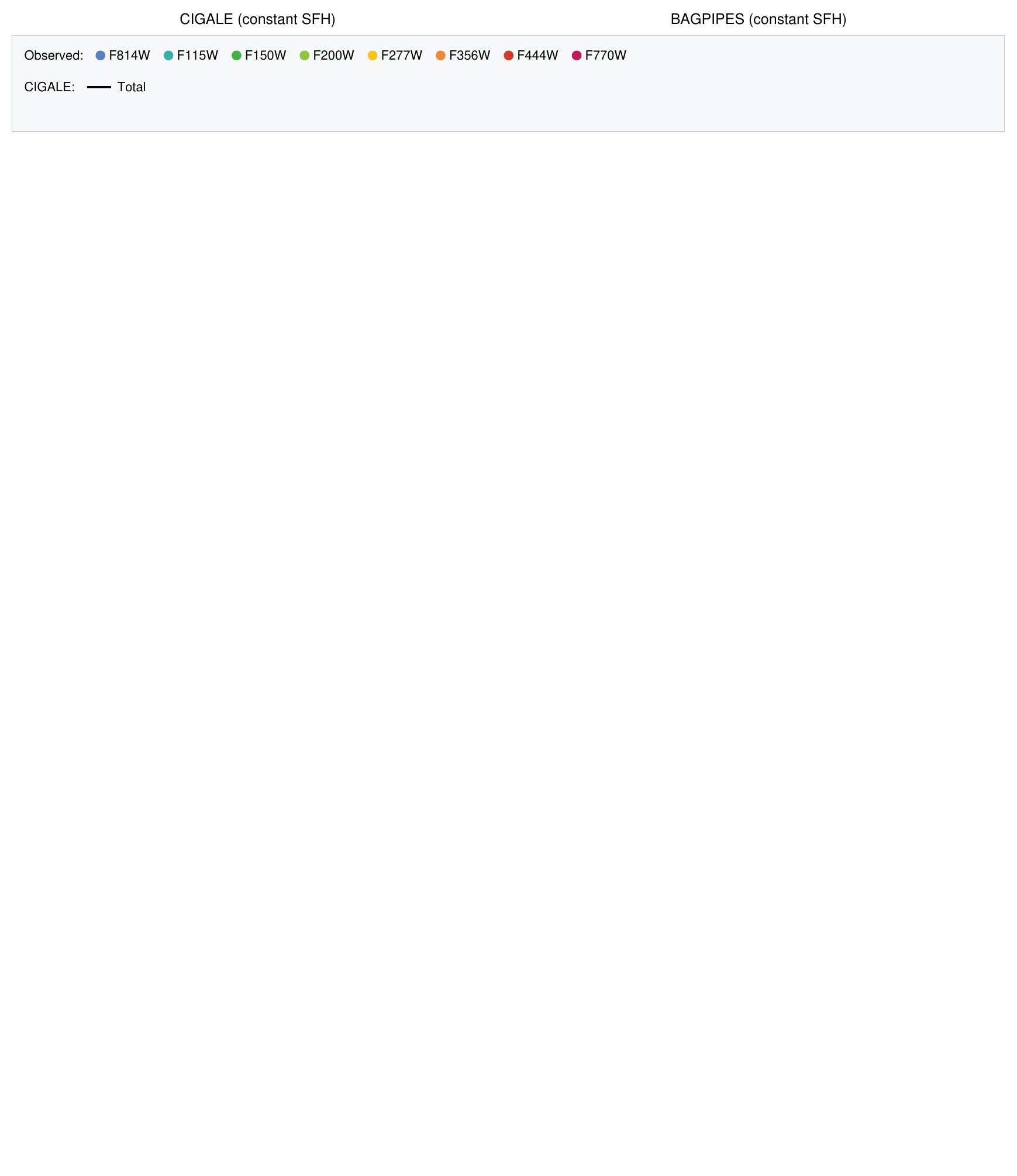}
    \caption{\texttt{CIGALE} constant-SFH fits to the companion photometry, sheet 1 of 6. Source identifiers are shown in the panels.}
    \label{fig:sed_fits}
\end{figure*}

\begin{figure*}[p]
    \centering
    \includegraphics[width=\textwidth,height=0.88\textheight,keepaspectratio]{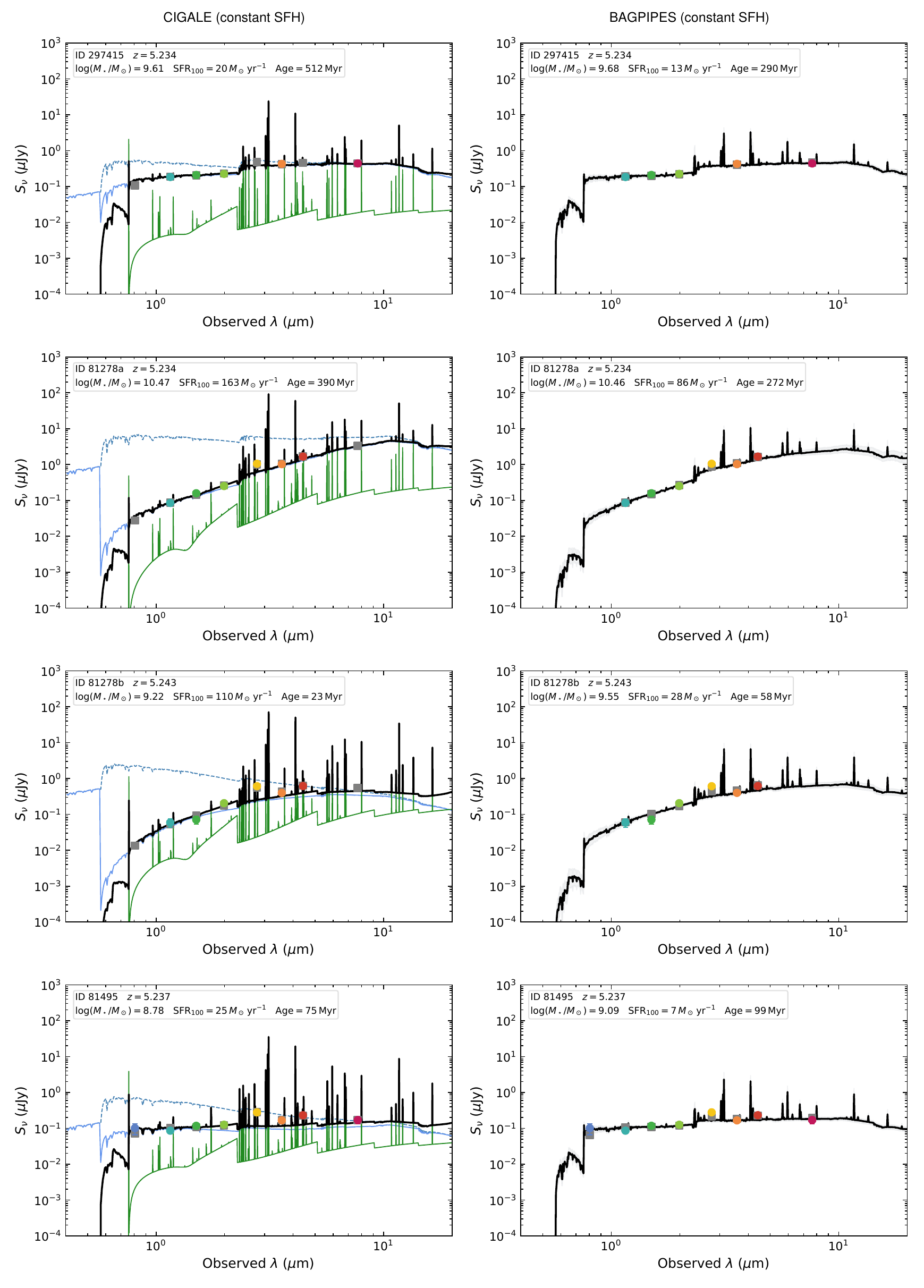}
    \caption{\texttt{CIGALE} constant-SFH fits to the companion photometry, sheet 2 of 6. Source identifiers are shown in the panels.}
    \label{fig:sed_fits_02}
\end{figure*}

\begin{figure*}[p]
    \centering
    \includegraphics[width=\textwidth,height=0.88\textheight,keepaspectratio]{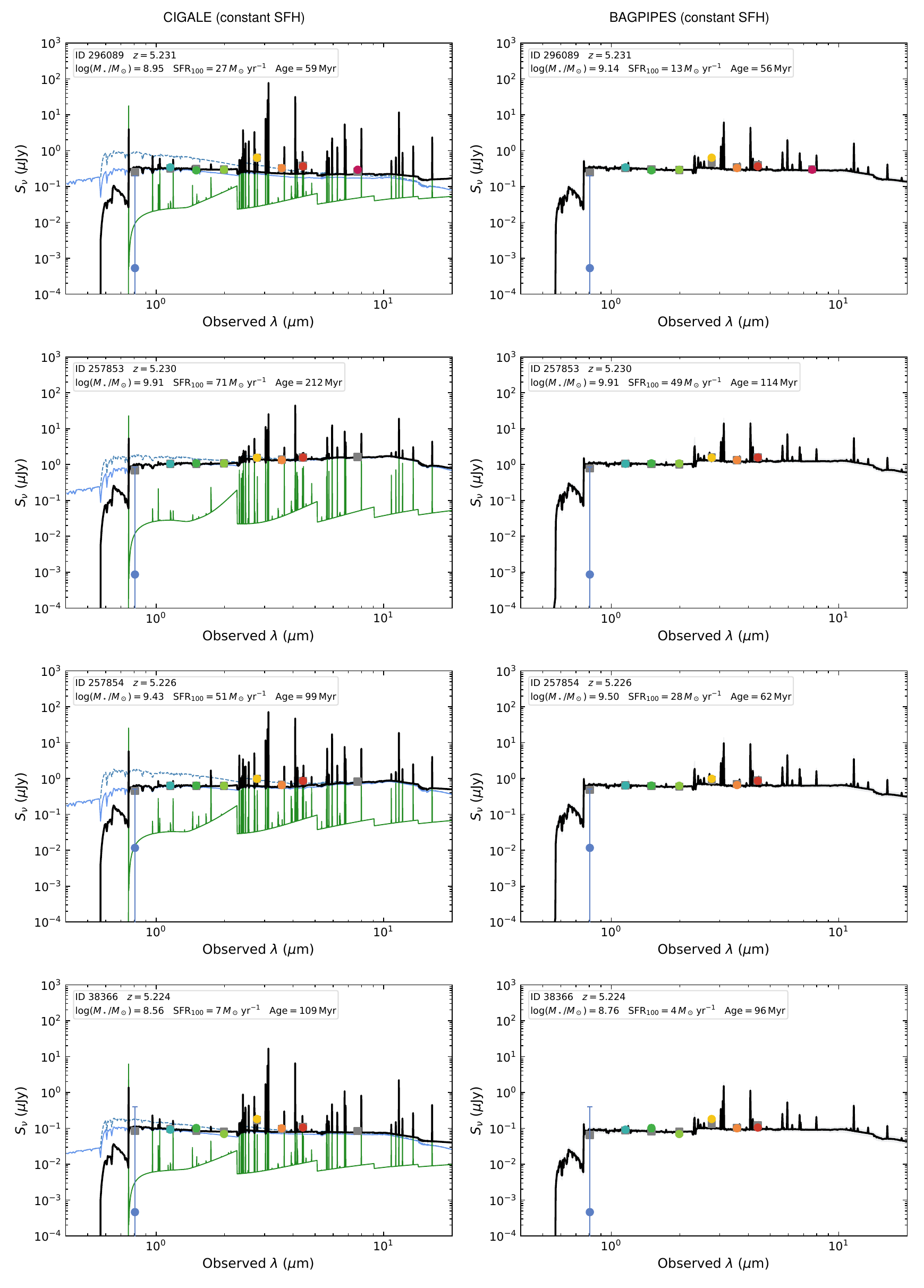}
    \caption{\texttt{CIGALE} constant-SFH fits to the companion photometry, sheet 3 of 6. Source identifiers are shown in the panels.}
    \label{fig:sed_fits_03}
\end{figure*}

\begin{figure*}[p]
    \centering
    \includegraphics[width=\textwidth,height=0.88\textheight,keepaspectratio]{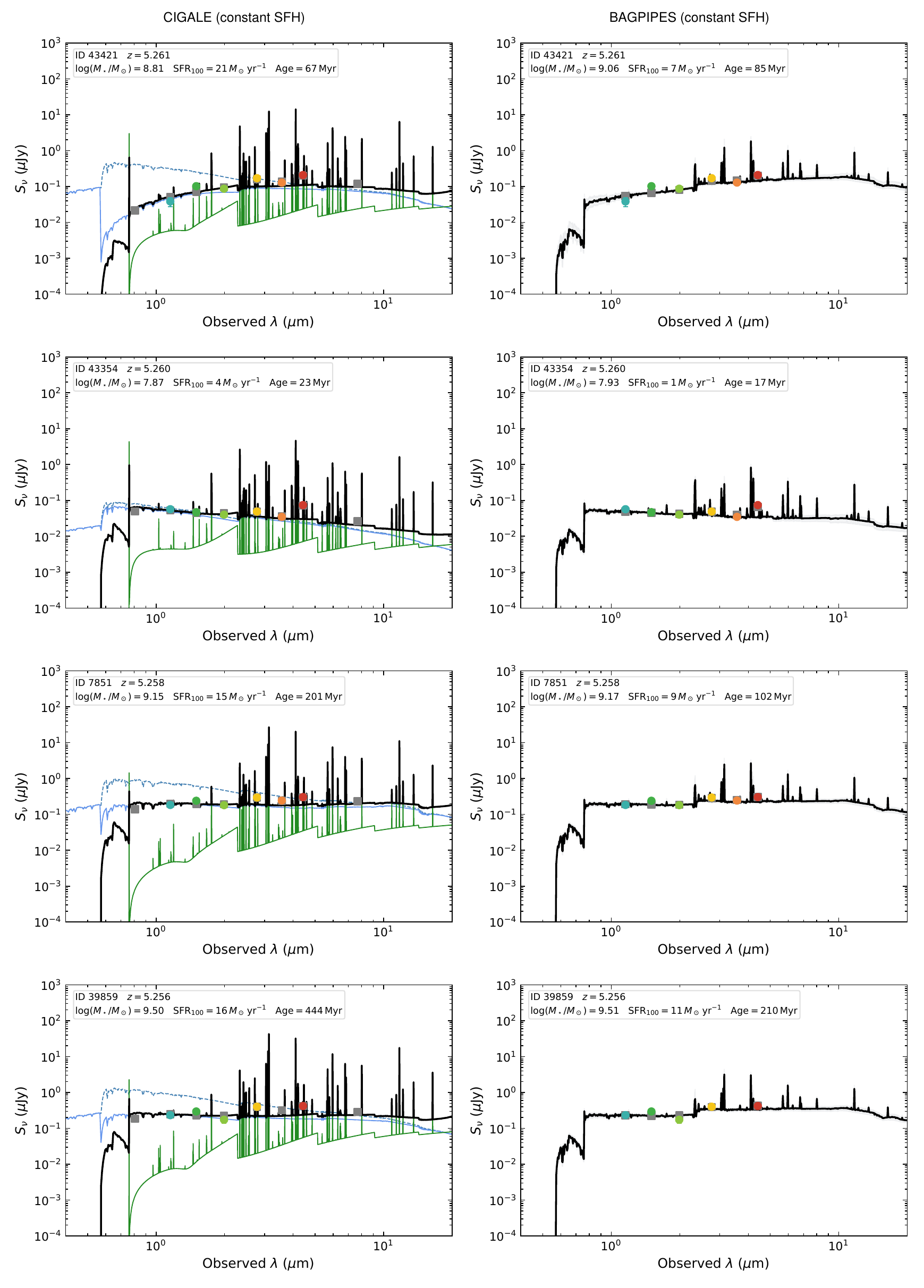}
    \caption{\texttt{CIGALE} constant-SFH fits to the companion photometry, sheet 4 of 6. Source identifiers are shown in the panels.}
    \label{fig:sed_fits_04}
\end{figure*}

\begin{figure*}[p]
    \centering
    \includegraphics[width=\textwidth,height=0.88\textheight,keepaspectratio]{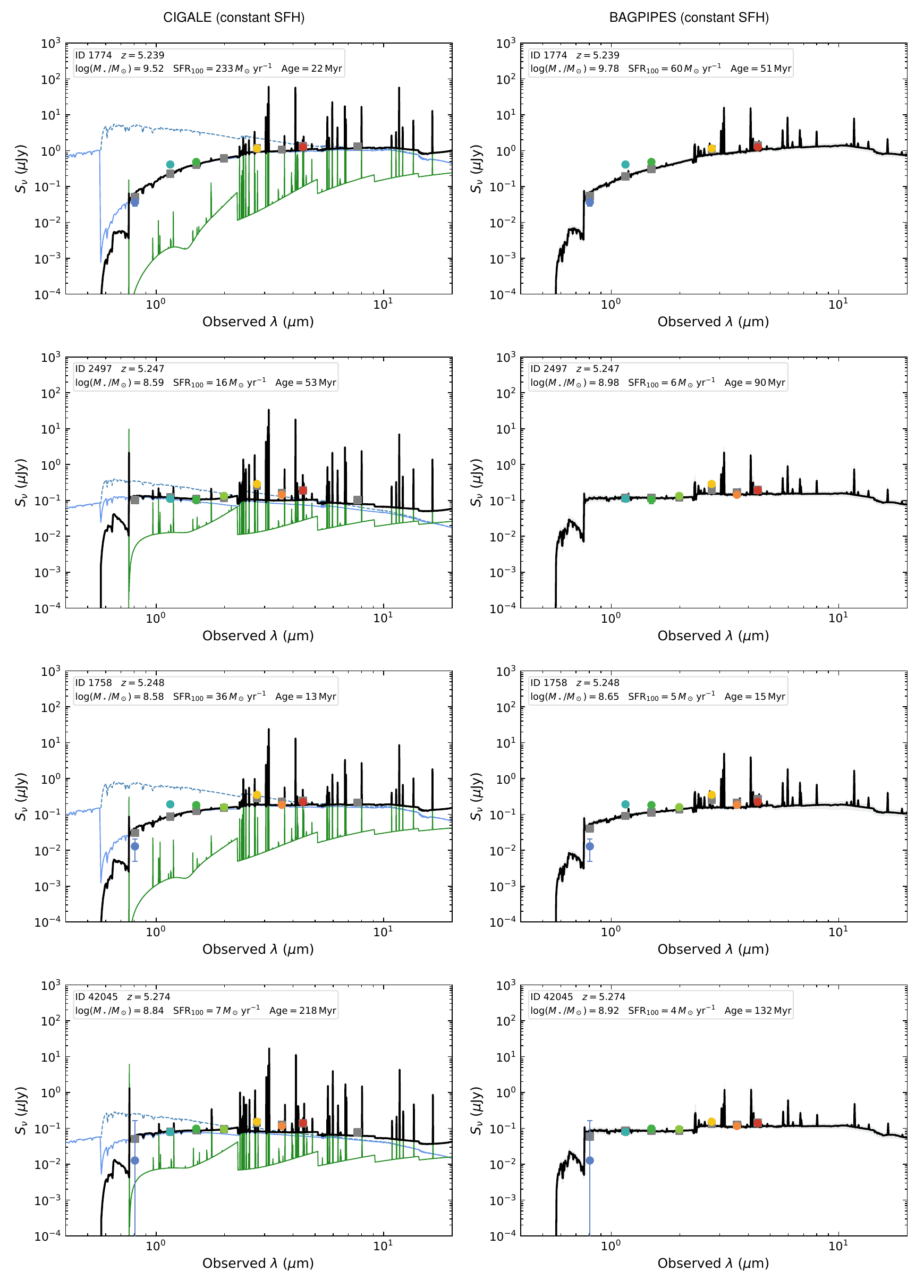}
    \caption{\texttt{CIGALE} constant-SFH fits to the companion photometry, sheet 5 of 6. Source identifiers are shown in the panels.}
    \label{fig:sed_fits_05}
\end{figure*}

\begin{figure*}[p]
    \centering
    \includegraphics[width=\textwidth,height=0.88\textheight,keepaspectratio]{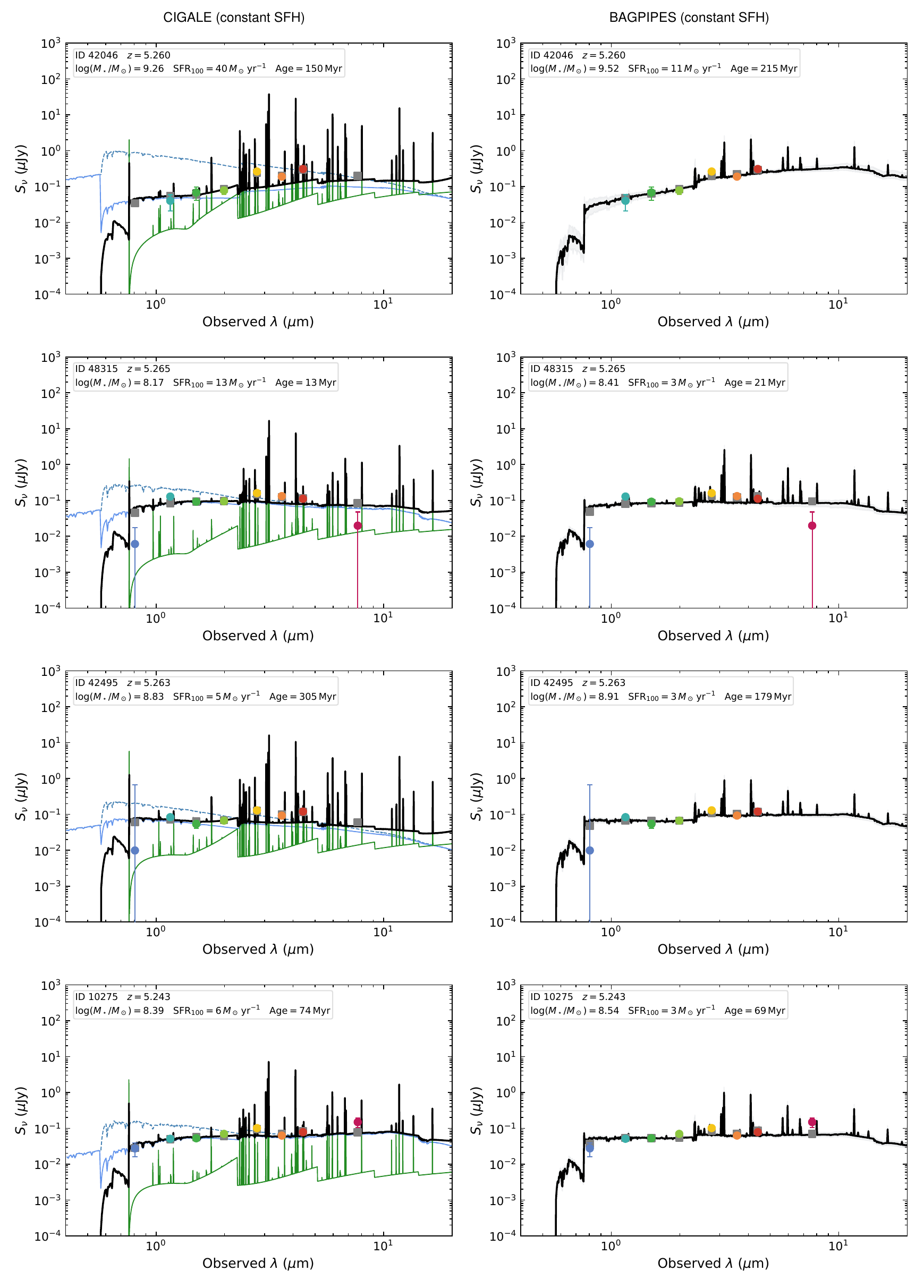}
    \caption{\texttt{CIGALE} constant-SFH fits to the companion photometry, sheet 6 of 6. Source identifiers are shown in the panels.}
    \label{fig:sed_fits_06}
\end{figure*}

Figures~\ref{fig:sed_fits}--\ref{fig:sed_fits_06} show the \texttt{CIGALE} constant-SFH fits to all companion components.

\IfFileExists{aasjournal7.1.bst}{%
  \bibliographystyle{aasjournal7.1}%
}{%
  \bibliographystyle{aasjournalv7.1}%
}
\bibliography{sample702}

\end{document}